\documentstyle[11pt,newpasp,twoside,epsf]{article}                          
\begin{document}
\title{Pulsar Death at an Advanced Age}

\author{Jonathan Arons}
\affil{Department of Astronomy, Department of Physics and 
Theoretical Astrophysics Center, University of California at 
Berkeley, 601 Campbell Hall, Berkeley, CA 94720-3411, USA  
arons@astroplasma.berkeley.edu} 


\begin{abstract}
I summarize the theory of acceleration of non-neutral particle 
beams by starvation electric fields along the polar magnetic field
lines of rotation powered pulsars, including the effect of dragging
of inertial frames which dominates the acceleration of a space charge
limited beam. I apply these results to a new calculation
of the radio pulsar death line, under the hypotheses that pulsar
``death'' corresponds to cessation of pair creation over the 
magnetic poles {\it and} that the magnetic field has a locally
dipolar topology.  The frame dragging effect in star centered
dipole geometry does improve
comparison of the theory with observation, but an unacceptably
large conflict between observation and theory still persists.
Offsetting the dipole improves 
the comparison, but a fully satisfactory
theory requires incorporating magnetic conversion of inverse Compton
gamma rays, created by scattering thermal photons from the surface of
old neutron stars ($t > 10^8 $ years) kept warm ($T \geq 10^5$ K)
by friction between the rotating core and the crust. The result is 
a ``death valley'' for pulsars; 
offsets of the dipole center from the stellar center in the oldest
stars $\sim (0.7-0.8) R_*$ suffice. The resulting theory
predicted the existence of rotation powered pulsars with these advanced
ages, a prediction confirmed by the recent discovery that PSR J2144-3933
actually has a rotation period of 8.5 seconds.
\end{abstract}

\section{Introduction}

Radio emission from Rotation Powered Pulsars (RPPs)
probably has its origin in the relativistic outflow
of electron-positron pairs along the polar magnetic field lines of
a dipole magnetic field frozen into the rotating neutron star (e.g., 
Arons 1992, Meszaros 1992). 
The evidence for dipole magnetic fields comes primarily
from the electromagnetic theory of RPP spindown energy losses, which occur
at the rate 
$
\dot{E}_R = k \mu^2 \Omega_*^4 /c^3 = 
   -I \Omega_* \dot{\Omega}_*
$
(Dyson 1971, Arons 1979, 1992).
Here $\mu$ is the magnetic moment, $\Omega_* = 2\pi /P$, $P$ is the 
rotation period, and $k$ is a
function of any other parameters of significance, with magnitude on
the order of unity.  In the vacuum theory (Deutsch 1955), 
$k =(2/3) \sin^2 i$, with $i$ the angle between the magnetic moment
and the angular velocity. Theoretical work on the
torques due to conduction currents steming back to Goldreich and Julian
(1969), coupled to the approximate independence of spindown torques from
observationally estimated values of $i$ (Lyne and
Manchester 1988), suggest that in reality $k$ does not substantially
depend on $i$. In the subsequent discussion, 
I assume $k = 4/9$, the average of the vacuum value over the sphere. 
Application of
this EM energy loss rate to the observations of RPPs' periods
and period derivatives yields $\mu \sim 10^{30}$ cgs for ``normal''
RPPs, and $\mu \sim 10^{27}$ cgs for millisecond RPPs. 

The electromagnetic torque interpretation of pulsar spindown
constrains only the exterior dipole moment of the magnetic field.
However, Rankin (1990) and Kramer {\it et al.} (1998) have presented 
strong evidence in favor of a low altitude ($r \approx R_*$) dipole 
geometry for the site of the core component of pulsar radio emission.
Arons (1993) gave evidence that spun up millisecond
pulsars must have substantially dipolar fields at low altitute.
  
If electron-positron pair creation above the polar caps is important for
radio emission, all observed pulsars must lie in the region of 
$P-\dot{P}$ space where polar cap acceleration has sufficient vigor 
to lead to copious pair production. Yet, to date, all 
{\it internally consistent} theories of polar cap pair 
creation have required hypothesizing a large scale ({\it e.g.}, quadrupole) 
component of the magnetic field with strength comparable to that of the
dipole (Ruderman and Sutherland 1975, Arons and
Scharlemann 1979, Barnard and Arons 1982, Gurevich and Istomin 1985).
Such strong magnetic 
anomalies contradict the evidence in favor of an apparently
dipolar low altitude geometry; the alteration of the magnetic geometry
also ruins the internal consistency of many models' electrodynamics. 

Here I describe a low altitude polar cap
acceleration theory which successfully associates pulsar ``death'' with
the cessation of pair creation in an {\it offset} dipole low altitude 
magnetic field.  The basic acceleration physics is that
of a space charge limited relativistic particle beam accelerated 
along the field lines by the starvation electric field,
as in the Arons and Scharlemann theory, but with the additional
effect of inertial frame dragging, first pointed out by Muslimov and Tsygan 
(1990, 1992) and by Beskin (1990). 
If the dipole's center is offset from the stellar center, the magnetic field at one pole becomes substantially
stronger than it would be if the same magnetic dipole were star centered.
The location of an individual pulsar's pair creation death depends
on the magnitude of the offset, thus yielding a ``death valley'' (Chen and
Ruderman 1993) for the whole pulsar population. Finally, if thermal photon
emission at temperature $T \sim 10^5$ K continues to great age 
($t > 10^8$ years), as is is expected in neutron star models with late time heating due to friction between the crust and core, the theory with dipole offsets predicts and accounts for pulsars with very long periods and great age. 

\section{Polar Acceleration}

Study of polar cap
relativstic particle acceleration in the 1970's had led to the 
conclusion that acceleration of a space charge limited particle beam from
the stellar surface with energy/particle high enough to emit
magnetically convertible curvature gamma rays occurs because of curvature 
of the magnetic field (Scharlemann {\it et al.} 1978, Arons and Scharlemann
1979). In a curved $B$ field, matching of the beam charge
density to the Goldreich-Julian density occurs only at the surface. 
Along field lines 
which curve toward the rotation axis (``favorably curved'' field lines), 
the beam fails to short out the vacuum above the surface, 
Therefore, particles accelerate along $B$ through a potential drop
$ \Delta \Phi_\parallel =\Delta \Phi_{SAF} \approx \Phi_{\rm pole} 
   ( R_*/\rho_B )  \sim 10^{-2} P^{-1/2} \Phi_{\rm pole}.$ 
The numerical value assumes field lines have dipolar radius of curvature
$\rho_B \sim \sqrt{R_* c/\Omega_*} $. 
Here
$
\Phi_{\rm pole} \equiv \Omega_*^2 \mu/c^2 = 
    1.09 \times 10^{13} (I_{45}/k)^{1/2}
      (\dot{P}_{15}/P^3)^{1/2}
$
Volts, with $\dot{P}_{15} \equiv \dot{P} /10^{-15} \; {\rm s/s}$ and 
$I_{45} = I/10^{45}$ g-cm$^2$. Particles drop through the potential 
$\Delta \Phi_{SAF}$ over a length $L_\parallel \sim R_*$. 

\begin{figure}
\plottwo{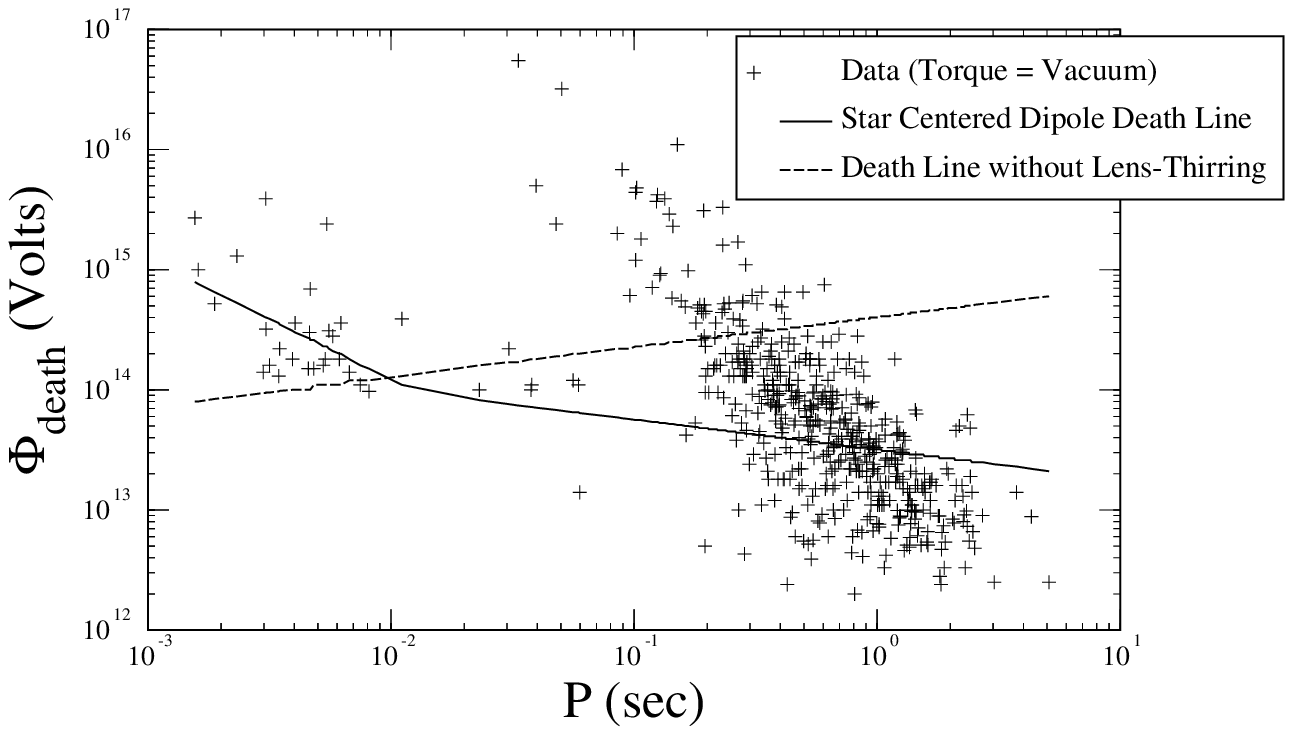}{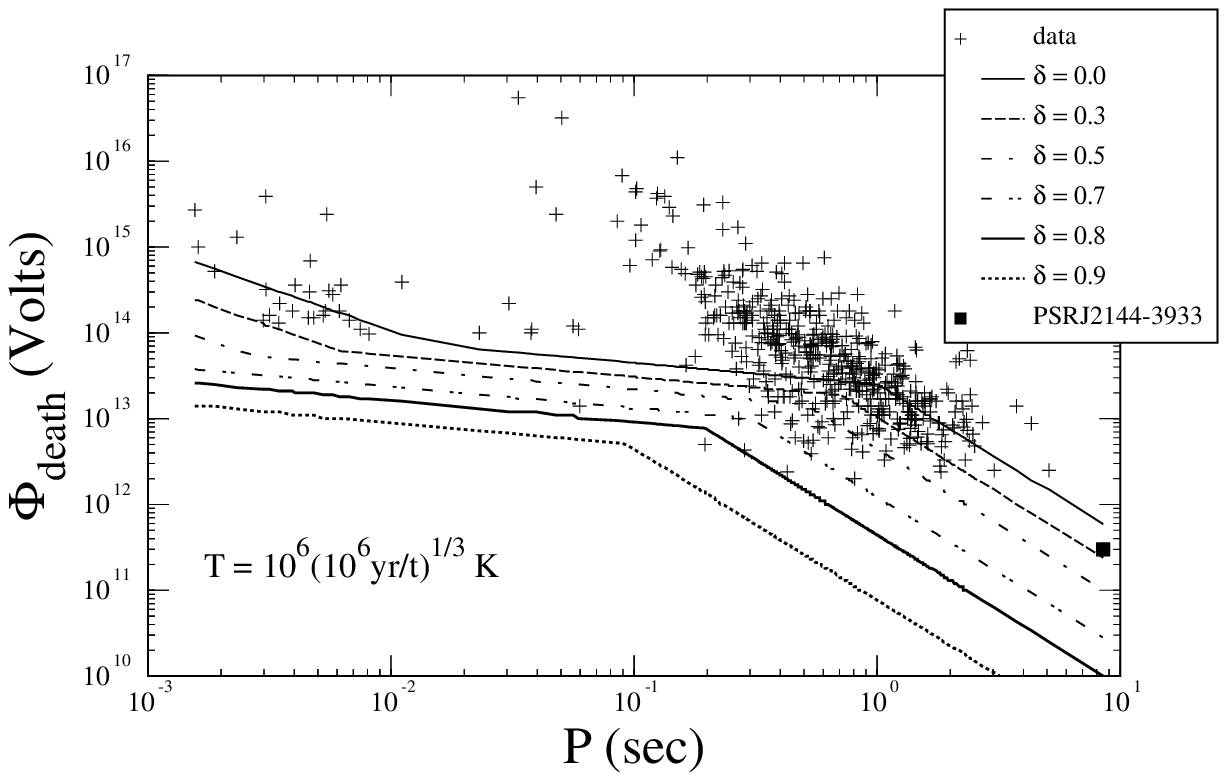}
\caption{Left panel: Pair creation Death Lines for Star Centered Dipoles.  The solid line
gives the result for curvature gamma ray emission and and one photon magnetic
conversion, when
the beam acceleration model incorporates the effect of inertial frame dragging.
The dashed line show the result for the same geometry and the same
gamma ray emission and absorption physics, but with inertial
frame dragging neglected in the particle acceleration theory.
Right panel: Death Valley for an offset vector parallel to the dipole moment,
  with both curvature and {\it resonant} inverse Compton emission as
  gamma ray sources, and with stellar temperature
  kept high by internal heating.  The filled square shows the 
  location of the recently discovered long period pulsar
  J2144-3933.}
\label{fig:centered}
\end{figure}

Curvature gamma rays have typical energy 
$\varepsilon_c \sim (\hbar c / \rho_B ) (e \Delta \Phi_\parallel / m c^2 )^3 
 \propto  \Phi_{\rm pole}^3 /\rho_B^4$, while the optical depth for pair
creation, due to one photon conversion of gamma rays emitted by electrons 
(or positrons) can be shown to be 
$\tau = \Lambda \exp[-a (mc^2 /\varepsilon_c) (B_q /B_*) (\rho_B /R_* )]$
(Arons and Scharlemann 1979, Luo 1996, Bjornsson 1996),
where $a$ is a pure number (typically  $ \sim 30$) and $\Lambda$ is a 
combination of the basic parameters which is quite large 
($\ln \Lambda \sim 20$). A reasonable
theoretical definition of the death line is $\tau = 1$. 
Using  $B_* = 2 (\Phi_{\rm pole} / R_*) (c / \Omega_* R_*)^2$, $\Delta \Phi_{SAF} $ 
and setting $\tau$ equal
to unity yields the death line, expressed as $\Phi_{\rm death} (P)$
such that stars  $\Phi_{\rm pole} < \Phi_{\rm death}$ do not make pairs.
This death line, appears as the dashed line in
Figure \ref{fig:centered}.  
This figure shows clearly that the large dynamic range in $\Phi_{pole}, \; P$ space made 
available by the cataloging of millisecond pulsars falsifies
this theory, even if one invokes non-dipolar radii of curvature to move the
position of the death line vertically in the diagram - the scaling with period
flatly disagrees with the shape of
the boundary of pulsar radio emission in the $\Phi_{pole}, \; P$ diagram.
These results imply either that something else governs the
low altitude acceleration which leads to pair creation, or that pair creation is
not important to radio emission.

Muslimov and Tsygan (1990, 1992) uncovered a
previously overlooked effect on the acceleration of the non-neutral beam from
the stellar surface. Stellar rotation drags the inertial frame into rotation,
at the angular velocity 
$\omega_{LT} = (2 G I/R_*^3 c^2) \Omega_* (R_* /r)^3$. Therefore, the
electric field required to bring a charged particle into corotation is
${\bf E}_{co}=-(1/c)[({\bf \Omega}_* - {\bf \omega}_{LT}) \times {\bf r}] 
\times {\bf B}$;
the magnetic field rotates with respect to {\it local} inertial space, not
inertial space at infinity. The charge density required to
support this {\it local} corotation electric field is 
$
\eta_R = - [({\bf \Omega}_* - {\bf \omega}_{LT}) \cdot {\bf B}]/ 2 \pi c =
  - [{\bf \Omega}_* \cdot {\bf B}/ 2 \pi c] 
    [ 1 - \kappa_g (R_* /r)^3 ],
$
where $\kappa_g = 2 G I /R_*^3 c^2 = 0.17 (I_{45} /R_{10}^3)$. Relativistic
space charge limited flow from the surface 
has a beam charge density 
$\eta_b  
 = - ({\bf \Omega}_* \cdot {\bf B}_* /2\pi c ) (1-\kappa_g) (B/B_*)$. Above the
surface, this charge density is too small to short out $E_\parallel$ on
{\it all} polar field lines, not just the favorably curved part of a polar
flux tube, thus providing a theoretical basis for polar cap acceleration models
to be in accord with the observed rough symmetry of radio emission 
with respect to
the magnetic axis (Lyne and Manchester 1988). One can graphically
describe this general relativistic origin of electrical starvation as
the consequence of the field lines rotating faster with respect to inertial
space as the radius increases, at the angular speed 
$\Omega_* - \omega_{LT}(r) = \Omega_* [1 - \kappa_g (R_* /r)^3 ]$.
The constraint of relativistic flow along
$B$ allows the beam to provide only a charge density sufficient to support 
corotation at the angular speed $\Omega_* (1 - \kappa_g)$. The difference
not surprisingly leads to an accelerating potential drop 
$
\Delta \Phi_\parallel \approx \kappa_g \Phi_{pole} [1 - (R_*/r)^3].
$
For normal ($P \sim 1$ sec)
pulsars with dipole fields,  
the effect of dragging of inertial frames on the beam's acceleration
yields curvature gamma ray energies 1000 times 
greater than occur in the Arons and Scharlemann pair creation theory; 
for MSPs, the theories yield 
comparable results, although of course the symmetry of the beam 
with respect to the magnetic axis differs.

\section{Death Lines and Death Valley}

When curvature emission is the only source
of gamma rays, the death line for a star
centered dipole is shown in Figure \ref{fig:centered}.
Dragging of inertial frames clearly improves the agreement between the
boundary of pair activity in the $\Phi - P$ diagram and the region where
pulsars occur, but the discrepancy is still too large - something else
is missing.  If the field geometry must be {\it locally} dipolar at low altitude, 
then the only ingredients still not included are 1) offset of the dipole 
from the stellar center and 2) additional gamma ray emission and absorption 
processes.
The simplest dipole offset has the magnetic
field of a point dipole, with the center of the dipole displaced
from the stellar center by an offset vector  parallel to 
${\bf \mu}$. This has the effect of increasing the
magnetic field at one pole to strength $B_* = 2 \mu / (R_* - \delta )^3$, with
a resulting drastic increase in the gamma ray opacity, while leaving the 
accelerating potential unaltered. The results with curvature radiation
as the only gamma ray emission process show that dipole offsets
do allow such a curvature radiation theory of pulsar death to survive the challenge
of modern observations, although at the price of displacements of the 
dipole center from the stellar center comparable to moving the dipole's 
center to the base of  the crust. Such a model still does not account for
PSR J2144-3933 (Young {\it et al.} 1999), however. 

Curvature emission is not the only means of converting beam energy to gamma rays.
ROSAT observations have revealed the long sought thermal X-rays from neutron star
surfaces (Becker and Tr\"{u}mper 1997).  Resonant Compton scattering creates 
magnetically convertible gamma rays at a spatial rate 
$(dN_\gamma /ds)_{rC}  \propto T /\Gamma^2$,
(e.g., Luo 1996) where $T$ is the temperature of the cooling 
neutron star (polar cap heating is unimportant in death valley) and 
$\Gamma = e \Phi /m_\pm c^2$ is the Lorentz factor of an electron or
positron in the beam. Compton scattering thus can become a significant
source of gamma rays in stars with {\it small} accelerating potentials. 
Compton scattering thus may contribute significantly for stars with low 
overall voltage. 

This expectation is correct,
{\it if} internal heating ({\it e.g.}, Umeda {\it et al.} 1993) keeps the 
surface temperature above $10^5$ K  at spindown ages in excess of $10^{7.5}$
years. In this case, resonant Compton scattering of thermal photons 
by a polar electron beam {\it does}
extend death valley to include all the observed pulsars, with 
offsets required in the lowest voltage RPPs of order 60\% to 70\%, as is shown 
in the right panel of Figure \ref{fig:centered}.
Note that this theory predicted (Arons 1998) the existence of RPPs with large
periods ($P > 5$ seconds) and unusually low voltage $\Phi_{pole} < 10^{12.5}$
Volts [great age, $t = 170 (10^{12} V /\Phi_{pole})^2 (10^s /P)^2$ Myr.]

Indeed, recent observations (Young {\it et al.} 1999) have found a
a $P = 8.51$ second pulsar, PSR J2144-3933, which falls below the death line
for a star centered dipole (even including Compton scattering), but lies
comfortably within death valley, when both Compton scatering and dipole offsets
are included.

\section{Conclusion}

I have shown that polar pair creation based on acceleration
of a steadily flowing, space charged limited non-neutral beam in a
locally dipolar magnetic geometry at low altitude
is consistent with pulsar radio
emission throughout the $P - \dot{P}$ diagram, provided 1) the effect of dragging
of inertial frames is included in estimates of the starvation electric field;
2) the dipole center is strongly offset from the stellar center in older stars, perhaps
as much as $0.7-0.8 R_*$; and 3) inverse Compton emission of thermal
photons from a neutron star cooling slower than exponentially at ages
in excess of $10^6$ years plays an important role in the emission of 
magnetically convertible gamma rays. Stars not deep in death valley have 
copious pair outflow on {\it all} polar field lines.  Such outflow shorts
out the ``outer gaps'' proposed as the dynamical basis for gamma ray 
emission in the outer magnetosphere (Cheng, Ho and Ruderman 1986, 
Romani 1996). Either polar cap gamma ray emission ({\it e.g.} Zhang
and Harding 2000) or possible outer magnetsophere emission from a 
dissipative return current layer (Arons 1981) remain as candidates for the
gamma ray emission observed in RPPs. 

 The development of new diagnostics
of the low altitude magneic field, gamma ray observations sensitive
to low altitude emission, and optical and UV observations of thermal 
emission from old, nearby RPPs will eventually provide tests of these ideas.

\section{Acknowledgments}

My research on pulsars is supported in part by NSF grant AST 9528271 and
NASA grant NAG 5-3073, and in part by the generosity of California's taxpayers.

\end{document}